\documentclass[aps,prl,twocolumn,groupedaddress]{revtex4}
\usepackage{graphicx,color,subfigure,amsmath}
\begin{document}

\title{Jamming of semiflexible polymers}
\author{Robert S. Hoy}
\email{rshoy@usf.edu}
\affiliation{Department of Physics, University of South Florida, Tampa, FL, 33620}
\date{\today}

\begin{abstract}
We study jamming in model freely rotating polymers as a function of chain length $N$ and bond angle $\theta_0$.  The volume fraction at jamming, $\phi_J(\theta_0)$, is minimal for rigid-rod-like chains ($\theta_0 = 0$), and increases monotonically with increasing $\theta_0 \leq \pi/2$. 
In contrast to flexible polymers, marginally jammed states of freely rotating polymers are highly hypostatic, even when bond and angle constraints are accounted for.  
Large aspect ratio (small $\theta_0$) chains behave comparably to stiff fibers: resistance to large-scale bending plays a major role in their jamming phenomenology.
Low aspect ratio (large $\theta_0$) chains behave more like flexible polymers, but still jam at much lower densities due to the presence of frozen-in 3-body correlations corresponding to the fixed bond angles.  
Long-chain systems jam at lower $\phi$ and are more hypostatic at jamming than short-chain systems.
Implications of these findings for polymer solidification are discussed.
\end{abstract}
\maketitle

Experimental investigation of the role that chemically specific, microscale interactions play in controlling the (generally nonequilibrium) multiscale structure of synthetic polymer solids - with the aim of establishing predictive relationships - is very difficult.
\textit{Colloidal} and \textit{granular} polymers (CGPs) -- chains of linked, macroscopic monomers \cite{miskin13,miskin14,zou09,sacanna10,brown2012,blaaderen2012,feng13} -- offer a promising alternative for studies of the packing and phase behavior of these systems.
Their larger size allows far easier observation of their structure on scales ranging from monomers to the bulk, using optical microscopy or even the naked eye \cite{zou09}.
More fundamentally, like their microscopic counterparts, their structural characteristics depend on factors such as chain topology (connectivity), monomer shape, and chain stiffness.
For example, the characteristic ratio $C_{\infty}$ \cite{flory53} is a controllable parameter in both microscopic synthetic polymers and CGPs \cite{zou09,blaaderen2012,feng13}, and is closely analogous to the aspect ratio $\alpha$ of convex anisotropic shapes such as ellipsoids, rods, and spherocylinders. 
Varying $C_{\infty}$ can naturally be expected to profoundly affect CGPs' bulk morphologies.
However, experimental study of CGPs remains in its infancy.
Only a few systems have been synthesized, and the factors affecting their packing at both the monomer and chain scales remain poorly explored.
In particular, simulation studies of CGP solidification have focused on flexible chains \cite{karayiannis08,karayiannis08b,foteinopoulou08,karayiannis09,karayiannis09b,karayiannis10,karayiannis10b,reichhardt11,karayiannis13b,ni13,hoy13}.
Thus there is a great need to characterize how parameters such as $C_{\infty}$ affect the morphologies of dense CGP phases.

In this paper, we study jamming of model freely rotating (FR) polymers \cite{flory53} composed of tangent spheres with fixed bond lengths ($\ell=\ell_0$) and bond angles ($\theta=\theta_0$).
Unlike those of flexible polymers, packings of FR polymers necessarily possess extensive 3-body intrachain-structural correlations arising from the fixed bond angles.
We show that these correlations produce profoundly different jamming phenomenology compared to that of flexible polymers that lack the $\theta=\theta_0$ constraint but are otherwise identical \cite{karayiannis08,karayiannis08b,foteinopoulou08,karayiannis09,karayiannis09b,karayiannis10,karayiannis10b,reichhardt11,karayiannis13b,ni13,hoy13}.

Our systems are composed of $N_{ch}$ chains, each containing $N$ monomers of mass $m$.  
All monomers interact via a harmonic potential $U_{H}(r) = 10\epsilon (1 - r/\sigma)^2 \Theta(\sigma-r)$, where $\epsilon$ is the energy scale of the pair interactions, $\sigma$ is monomer diameter, and $\Theta$ is the Heaviside step function.
This purely repulsive interaction reflects the essentially athermal nature of real CGPs \cite{miskin13,miskin14,zou09,sacanna10,brown2012,blaaderen2012,feng13}.
Covalent bonds are modeled using the harmonic potential $U_c(\ell) = (k_c/2)(\ell-\sigma)^2$, leading to tangent-sphere polymers with equilibrium bond length $\ell_0=\sigma$.  
To closely approximate the fixed-length bonds of FR chains, we choose $k_c = 600\epsilon/\sigma^2$.
Angular interactions between three consecutive monomers along a chain are modeled by the harmonic potential $U_a(\theta) = (k_a/2)(\theta-\theta_0)^2$, where $\theta$ is the angle between consecutive bonds and is zero for straight trimers.  
FR chains are obtained in the limit $k_a \to \infty$; we choose $k_a = 600\epsilon/\rm{radians}^2$, which limits deviations from $\theta=\theta_0$ to less than $2^{\circ}$ for all conditions studied herein.
$C_{\infty} = (1 + cos(\theta_0))/(1 - cos(\theta_0))$ increases from $1$ to $\infty$ as $\theta_0$ decreases from $90^{\circ}$ to $0$.
We will contrast results for these systems to those for fully-flexible chains ($k_a = 0$), which have been extensively studied \cite{karayiannis08,karayiannis08b,foteinopoulou08,karayiannis09,karayiannis09b,karayiannis10,karayiannis10b,reichhardt11,karayiannis13b,ni13,hoy13} but not yet compared to semiflexible chains.

We prepare our systems using standard molecular dynamics techniques.
All MD simulations are performed using LAMMPS \cite{plimpton95}.
Initial states are generated by placing $N_{ch}$ FR chains
randomly within a cubic cell of side length $L$.  
Periodic boundary conditions are applied in all three directions.
The monomer number density is $\rho = N_{ch}N/L^3$, and the packing fraction is $\phi = \pi\rho\sigma^3/6$.
Newton's equations of motion are integrated with a timestep $\delta t = .005\tau$, where the unit of time is $\tau=\sqrt{m\sigma^2/\epsilon}$.
Temperature is maintained with a Langevin thermostat.
All systems are equilibrated at $k_BT/\epsilon = 1$ and $\phi = .03\phi_{xtal}$ (where $\phi_{xtal}=\pi/\sqrt{18}$ is the volume fraction of close-packed crystals) until both intrachain and interchain structure have converged.
The systems are then cooled to $T=0$ at a rate $10^{-5}/\tau$.

After cooling, jamming is simulated by compressing the systems very slowly.  
$L$ is varied in time as $L(t) = L_0 \exp(-\dot\epsilon t)$ until $\phi=\phi_{xtal}$.  
We choose $\dot\epsilon = 10^{-6}/\tau$, which is the slowest rate feasible for our employed system size ($N_{ch}N=20000$).
We have verified that strain rate dependence of all quantities measured herein is weak at this $\dot{\epsilon}$.
Jamming is defined to occur when the nonkinetic part of the pressure $P$ exceeds $P_{thres}=.01\epsilon/\sigma^3$ \cite{pthres}.
We choose to identify jamming with the emergence of a finite bulk modulus \cite{reichhardt11,vanHecke09} rather than with the vanishing of soft modes \cite{ohern02} because proper handling of soft modes associated with ``flippers'' (interior monomers with zero or one noncovalent contacts \cite{karayiannis09b}) is highly nontrivial.

\begin{figure}
\includegraphics[width=2.9in]{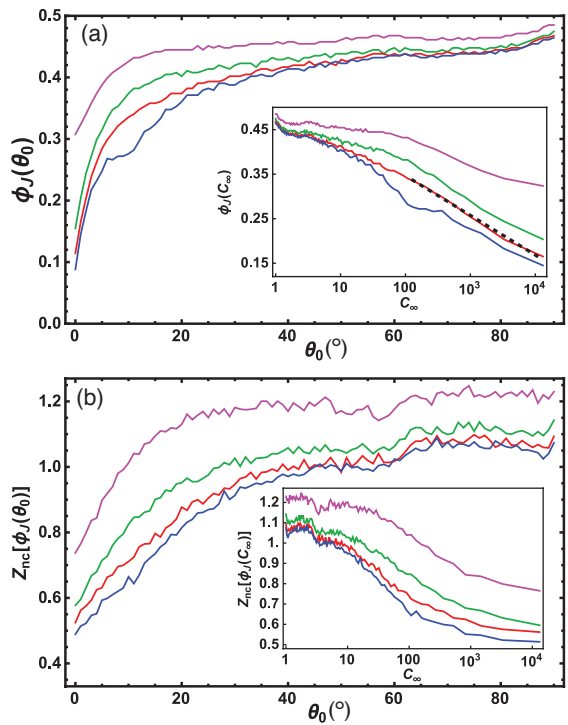}
\caption{Jamming for freely rotating polymers:\ $N$ and $\theta_0$ dependence.  Panel (a): $\phi_J(\theta_0)$ for chains of length $N=10$ (purple), $N=25$ (green), $N=50$ (red), and $N=100$ (blue).  The inset shows the same results plotted vs.\ $C_{\infty}$.  The dotted black line is a fit of the $N=50$ data to $\phi_J(C_{\infty}) = a - b\ln{(C_{\infty})}$.  Panel (b): $Z_{nc}[\phi = \phi_J(\theta_0)]$ for the same systems.}
\label{fig:phiJ}
\end{figure}

A polymer chain with spherical monomers and unconstrained covalent bonds and bond angles has $3N-6$ internal degrees of freedom (d.o.f.) - the same as any other collection of $N$ spherical particles.
Fixing the lengths of covalent bonds (imposing $\ell = \ell_0$) eliminates one d.o.f.\ per bond, or $N-1$ d.o.f.\ per chain.
Fixing bond angles (imposing $\theta = \theta_0$) eliminates one d.o.f per angle, or $N-2$ d.o.f.\ per chain.
Thus ideal freely rotating chains \cite{flory53} have $\mathcal{N}_{dof} = (3N-6)-(N-1)-(N-2) = N-3$ internal d.o.f.\ per chain.
Not coincidentally, $N-3$ is also the number of dihedral angles $\psi$ per chain, and chain configurations (modulo rigid translations/rotations) can be fully described by the values $\psi_i$ for $i=1,2,...,N-3$.

Isostaticity for freely rotating polymers occurs when the average coordination number for \textit{noncovalent} contacts, $Z_{nc}$, satisfies $Z_{nc} = 2\mathcal{N}_{dof}/N$.
Only noncovalent contacts are considered, because the stiff covalent bonds are treated as constraints which reduce $\mathcal{N}_{dof}$.
Thus in the long-chain limit one expects jamming at $Z_{nc}=2$ \textit{if} freely rotating polymers jam via the same mechanisms as spherical particles \cite{ohern02} and flexible polymers \cite{karayiannis08,karayiannis09b,karayiannis10b,reichhardt11}.
If they jam via different mechanisms, more like those involved in the jamming of ellipsoids \cite{donev04,mailman09} or spherocylinders \cite{kyrylyuk11}, one might expect jamming at $Z_{nc} < 2$.

Figure \ref{fig:phiJ} shows results for $\phi_J(\theta_0)$ at jamming (panel (a)) and $Z_{nc}[\phi=\phi_J(\theta_0)]$ (panel (b)) for chains of length $10 \leq N \leq 100$.
$\phi_J$ is lowest for rodlike chains with $\theta_0=0^{\circ}$: $N=100$ systems jam at volume fractions below $0.1$.
This is expected since rods composed of tangent beads have been shown to pack anti-optimally, i.e.\ to minimize $\phi_J$ \cite{miskin14}.
As $\theta_0$ increases, $\phi_J$ increases smoothly \cite{nonsmooth} with a nearly monotonically decreasing rate $\partial \phi_J/\partial\theta_0$ up to $\theta_0 \simeq 75^{\circ}$, then increases slightly more rapidly over the range $80^{\circ} \lesssim \theta_0 \leq 90^{\circ}$.
The inset shows the same results plotted against $C_{\infty}$.
$\phi_J$ decreases monotonically with increasing $C_{\infty}$, a trend closely analogous to the decrease of $\phi_J$ for ellipsoids and other anisotropic rigid convex particles  with increasing $\alpha$ \cite{donev04,mailman09,kyrylyuk11}. 
For stiff chains ($\theta_0 \lesssim 10^{\circ};\ C_{\infty} \gtrsim 10^2$) this decrease is logarithmic in $C_{\infty}$; 
this is presumably related to the slow freezing out of the configurational freedom associated with dihedral angles as chains approach the rodlike limit  (as $\theta_0 \to 0$).

FR polymers jam at much lower $\phi$ than their flexible counterparts; maximal values of $\phi_J$ occur for $\theta_0 \simeq 90^{\circ}$ and are about $0.47$ for long chains.
In contrast, for the protocol employed here, fully flexible $N=50$ chains jam at $\phi_J^{flex}=.618$ \cite{phijmon}, which is close to the monomeric value $\phi_J^{mon} = .636$ \cite{ohern02}.
The small difference $\phi_J^{mon}-\phi_J^{flex} \simeq .02$ indicates that the large differences $\phi_J^{mon}-\phi_J(\theta_0)$ shown in Fig.\ \ref{fig:phiJ}(a) arise primarily from FR polymers' constrained bond angles.
These differences can arise from differences in local packing associated with the (approximate) $\theta=\theta_0$ constraints, or from chain-bending stresses associated with our finite-stiffness angular potential $U_a(\theta)$, or both.
Note that they are \textit{not} associated with the loop-formation that reduces $\phi_J$ in flexible granular polymers \cite{zou09,reichhardt11}; loop formation is strongly suppressed in FR polymers by the stiff angular interactions.

\begin{figure*}
\includegraphics[width=7.0in]{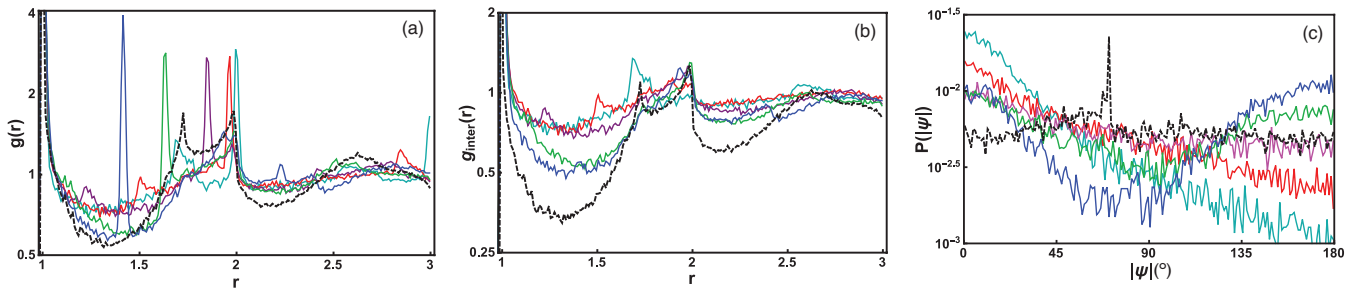}
\caption{Interchain and intrachain structure of marginally jammed states.  Solid curves show results for $N=50$ FR chains with $\theta_0=5^{\circ}$ (cyan), $\theta_0=23^{\circ}$ (red),  $\theta_0=45^{\circ}$ (purple),  $\theta_0=71^{\circ}$ (green), and $\theta_0=90^{\circ}$ (blue) at their respective $\phi_J(\theta_0)$.  Dashed black curves show results for fully-flexible $N=50$ chains at $\phi_J^{flex}$.  Panel (a):\ pair correlation function $g(r)$; panel (b):\ its interchain component $g_{inter}(r)$; panel (c):\ the dihedral angle distribution $P(|\Psi|)$.}
\label{fig:interintra}
\end{figure*}

Fully flexible polymers jam at isostaticity, i.e.\ at $Z_{nc}=4$ (when the average total coordination number $Z=6$) \cite{karayiannis08,karayiannis09b,karayiannis10b}. 
Figure \ref{fig:phiJ}(b) shows that the marginally jammed states at $\phi=\phi_J(\theta_0)$ are highly hypostatic.
$Z_{nc}[\phi=\phi_J(\theta_0)]$ increases monotonically with increasing $\theta_0$ (decreasing $C_{\infty}$), but even $\theta_0=90^{\circ}$ chains have $Z_{nc} < 1.1$ in the long-chain limit.
Stiff (low-$\theta_0$) chains are less free to avoid jamming by local dihedral rotations that move monomers on different chains away from one another, and hence jam at a greater degree of hypostaticity.
The free volume swept out by monomer $i$ as the dihedral angle $\psi_i$ varies from $-180^{\circ}$ to $180^{\circ}$ (with the other three monomers composing $\psi_i$ held fixed) is $V_{\psi} \simeq (\pi\sigma^2 \ell_0 / 4) \sin(\theta_0)$.
Thus the configurational freedom associated with the dihedral angles vanishes in the $\theta_0 \to 0$ limit, and the number of \textit{effective} d.o.f.\ per dihedral angle drops, with a corresponding drop in the number of contacts at jamming.
As we will show below, the source of this strongly hypostatic jamming is resistance to large-scale chain bending arising from the stiff angular interactions.

The chain length depencences of the above quantities are also illustrative.
As expected \cite{zou09,reichhardt11}, $\phi_J$ decreases with increasing $N$, scaling roughly as $\phi_J(N; \theta_0) = \phi_J(\infty; \theta_0) + c(\theta_0)/N$.
The strength of the chain length dependence ($c(\theta_0)$) is roughly constant for $\theta_0 \gtrsim 45^{\circ}$.
For straighter chains, $c(\theta_0)$ increases with decreasing $\theta_0$, reaching a maximum for rodlike chains with $\theta_0=0$; this is expected since $N/C_{\infty}$ decreases with decreasing $\theta_0$.
Shorter chains are less hypostatic for two reasons.
First, they jam at higher densities, which is naturally consistent with larger $Z_{nc}$.
Second, their shorter length makes them less prone to mechanical stabilization by long-range bending forces comparable to that observed in fibre networks \cite{broedersz11}.

Next we relate the above results to the local intrachain and interchain structure of marginally jammed states.
We have verified that systems remain isotropic; nematic order \cite{degennes79} remains minimal during compression for all $\theta_0$.
For the remainder of the paper we focus on $N=50$ systems, which are clearly in the long-chain limit \cite{nonsmooth}, particularly for $\theta_0 > 20^{\circ}$.
We will focus on five characteristic chain stiffnesses:\ rodlike chains with $\theta_0 = 5^{\circ}$ ($C_{\infty} \simeq 525$), intermediate-stiffness chains with $\theta_0 = 23^{\circ}$ and $45^{\circ}$ ($C_{\infty} \simeq 25\ \rm{and}\ 6$), alkane-like chains with $\theta_0 = 71^{\circ}$ ($C_{\infty} \simeq 2$), and low-aspect-ratio chains ($\theta_0 = 90^{\circ}$) that retain fixed bond angles but have $C_{\infty} = 1$.

Figure \ref{fig:interintra} shows the pair correlation function $g(r)$ (panel (a)), its interchain component $g_{inter}(r)$ (panel (b)), and the distribution of dihedral angles $P(|\psi|)$ (panel (c)) for FR chains of the five characteristic stiffnesses, as well as for $N=50$ fully-flexible chains, at their respective $\phi_J$.
All systems' $g(r)$ have a strong peak at $r=\sigma$ corresponding to both covalent and noncovalent contacts.
FR chains have strong second peaks at $r_{2nd} = 2\cos(\theta_0/2)\sigma$ that arise from the fixed bond angles.
The influence of these \textit{intrachain} constraints on local \textit{interchain} packing is strong.
The $\theta=\theta_0$ constraints strongly limit the ways in which the noncovalently bonded neighbors in the first coordination shell of a given monomer can arrange themselves, and therefore reduce $\phi_J$.
Moreover, FR chains do not collapse nearly as much during compression as fully-flexible chains, as shown by flexible chains' much lower values of $g_{inter}(r)$ in the range $1 < r/\sigma < 5/3$.
This means that on average, more chains are present in the immediate vicinity of a given monomer in a FR-polymer jammed state than in a fully-flexible-polymer jammed state.
This tighter interchain packing, in combination with the greater density of constraints, logically promotes jamming at lower $\phi_J$.

\begin{figure*}
\includegraphics[width=7.0in]{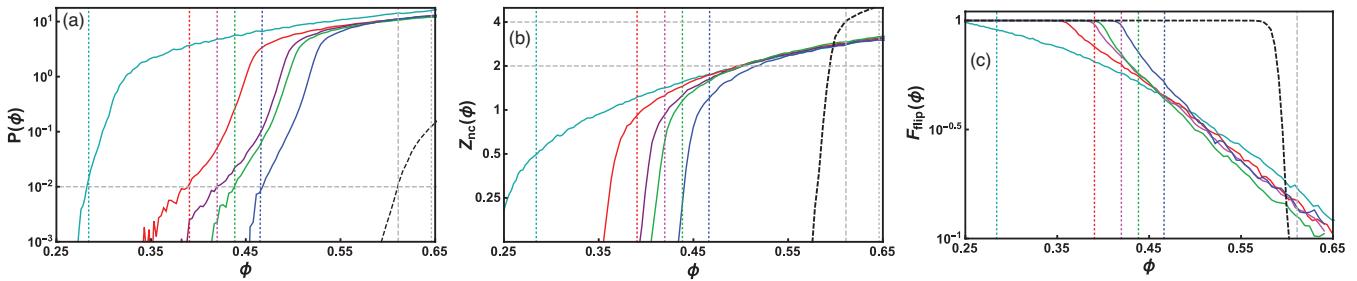}
\caption{
Dynamics of FR-polymer jamming: Pressure $P(\phi)$ (panel (a)), average number of noncovalent contacts per monomer $Z_{nc}(\phi)$ (panel (b)), and flipper fraction $F_{flip}(\phi)$ (panel (c)), for the same systems considered in Figure \ref{fig:interintra}.  Vertical dashed lines indicate $\phi_J$ for the corresponding systems, while the horizontal dashed lines indicate $P=.01\epsilon/\sigma^3$ (panel (a)) and $Z_{nc}=2$ and $4$ (panel (b)).} 
\label{fig:dynamics}
\end{figure*}

The fixed-angle constraints of FR polymers also dramatically influence their dihedral-angle distributions at jamming.
$P(|\psi|)$ for fully-flexible chains shows a sharp peak at $|\Psi| = 71^{\circ}$ associated with chain collapse into polytetrahedral local structure  \cite{karayiannis09b}, but is otherwise nearly flat \cite{polytet}.
FR-polymer chains do not collapse into polytetrahedral structures -- a key distinction since the incommensurability of differently oriented tetrahedra is a key factor promoting jamming in flexible-polymeric and monomeric systems \cite{anikeenko07,karayiannis09b}.
High-$C_{\infty}$ chains show a large, broad peak at $\psi = 0$ corresponding to \textit{cis} conformers.
The excess of \textit{cis} conformers arises from chain segments bending into arcs with radii of curvature $\sim 360/[2\pi\theta_0(^\circ)]$ during compression.
This trend gradually weakens as $\theta_0$ increases.
For large $\theta_0 \gtrsim 75^{\circ}$, a second peak corresponding to \textit{trans} conformers ($|\psi|=180^{\circ}$) appears.
This peak strengthens with increasing $\theta_0$ until \textit{cis} and \textit{trans} conformers are nearly equally likely for $\theta_0=90^{\circ}$.
For this $\theta_0$, the local four-monomer chain sections corresponding to individual dihedrals show a strong preference for planar (\textit{cis} and \textit{trans}) as opposed to maximally-nonplanar ($|\psi|=90^{\circ}$) conformations.
No such effects are present in fully-flexible polymers \cite{karayiannis09b}.

Next we examine the dynamics during compression, to better understand how the bond-angle constraints affect jamming as a \textit{process}.
Figure \ref{fig:dynamics} shows the pressure $P(\phi)$ (panel (a)), $Z_{nc}(\phi)$ (panel (b)), and the fraction of flippers $F_{flip}(\phi)$ (panel (c)), for the same six systems considered in Fig.\ \ref{fig:interintra}.
For fully-flexible polymers, $P(\phi)$ is concave-down at all $\phi \gtrsim \phi_J^{flex}$.
FR polymers show a qualitatively different behavior:\ $P(\phi)$ is concave-down at low $\phi$, then concave-up at intermediate $\phi$, then concave-down again at large $\phi$.
The first inflection point occurs at $\phi \simeq \phi_J$ and corresponds to the dominant contributor to $P$ shifting from pair and bond forces to angular forces arising from semiflexible chains' resistance to bending and collapse.
This shift is associated with stabilization against shear deformation (i.e., jamming) \cite{pthres,broedersz11} and supports our choice of $P_{thres}=.01\epsilon/\sigma^3$.
The second inflection point occurs well above $\phi_J$ and corresponds to a shift back to dominance of pair and bond contributions arising from strong intermonomer overlap.
This shift is why the $P(\phi)$ results for all but the stiffest systems fall onto a common curve at large $\phi$.

$Z_{nc}(\phi)$ increases much more gradually for FR polymers than for their fully flexible counterparts -- increasingly so as $\theta_0$ decreases.
The volume swept out by rigidly rotating chains (where the rotations are caused by interchain ``collisions'') increases with increasing $C_{\infty}$, in turn increasing the rate of such collisions and the associated pressure thereof.
Jamming occurs at (approximately) the $\phi$ that maximize $\partial^2 Z_{nc}/\partial \phi^2$.
These maxima seem to be associated with the abovementioned switch from pair/bond-dominated to chain-bending dominated pressure contributions.
In other words, the rate of increase of $Z_{nc}$ drops sharply as chains begin to interlock and bend.
Interestingly, the $Z_{nc}(\phi)$ results for different $\theta_0$ also fall onto a common curve at large $\phi$.
For $\phi \gtrsim \phi_J^{flex}$, $Z_{nc}(\phi)$ for FR polymers drops below its value for fully-flexible polymers because the stiff bond angles favor increasing pressure on existing interchain contacts over formation of new contacts.

As shown in Fig.\ \ref{fig:dynamics}(c), the fraction of unconstrained interior monomers (``flippers'' \cite{karayiannis09b}) drops much more gradually with increasing $\phi$ for FR chains than for fully-flexible chains.
The functional form of the drop is in $F_{flip}(\phi)$ is qualitatively different, and most critically, $F_{flip}(\phi)$ does not drop to near zero at $\phi_J$.
Indeed, a large fraction of interior monomers remain flippers (able to undergo dihedral rotations) at $\phi_J(\theta_0)$.
In other words, a large fraction of FR chains' internal d.o.f.\ remain unconstrained at jamming, consistent with the hypostaticity of marginally-jammed states discussed above.

These results suggest the following picture of semiflexible-polymer jamming:
First, collisions between chains during compression create local high-density regions that tend to capture chain segments within them.
Second, as compression continues, sections of chains between these captured segments collapse via dihedral rotations favoring \textit{cis} conformers; the extent of this collapse decreases with increasing $C_{\infty}$.
Third, once the dihedral degrees of freedom begin to exhaust, angular forces opposing chain-bending grow rapidly, producing jamming.

We emphasize that we have here considered dynamical jamming under a protocol that preserves chain uncrossability.
Other protocols such as those of Refs.\ \cite{karayiannis08, karayiannis08b,foteinopoulou08,karayiannis09,karayiannis09b,karayiannis10,karayiannis10b} will likely produce both higher $\phi_J$ and different $\theta_0$- and $N$-dependence; for example, FR chains with $\theta_0=0$ or $60^{\circ}$ can form close-packed crystals with $\phi=\phi_{xtal}$ at zero pressure.
However, we expect that our results will be directly relevant to experimental studies of CGPs since chain uncrossability is a critical feature of real polymeric systems.
Since the equilibrium bond angle (i.e., $\theta_0$) in experimental colloidal-polymer systems  can be controlled by grafting DNA binding ``patches'' onto emulsion droplets \cite{feng13}, producing systems that closely approximate FR polymers, our results should be directly applicable to future CGP experiments.
Moreover, since the decrease in $\phi_J$ with increasing $C_{\infty}$ reported here is closely analogous \cite{zou09} to the well-known increase in $T_g$ with increasing $C_{\infty}$ in microscopic synthetic polymers \cite{strobl07,ballauff89}, our results may be applicable to understanding the chain-stiffness dependence of the polymeric glass transition.

Helpful discussions with Corey S.\ O'Hern and Abram H.\ Clark, and support from NSF Grant No.\ DMR-1555242, are gratefully acknowledged.


\begin{thebibliography}{34}
\expandafter\ifx\csname natexlab\endcsname\relax\def\natexlab#1{#1}\fi
\expandafter\ifx\csname bibnamefont\endcsname\relax
  \def\bibnamefont#1{#1}\fi
\expandafter\ifx\csname bibfnamefont\endcsname\relax
  \def\bibfnamefont#1{#1}\fi
\expandafter\ifx\csname citenamefont\endcsname\relax
  \def\citenamefont#1{#1}\fi
\expandafter\ifx\csname url\endcsname\relax
  \def\url#1{\texttt{#1}}\fi
\expandafter\ifx\csname urlprefix\endcsname\relax\def\urlprefix{URL }\fi
\providecommand{\bibinfo}[2]{#2}
\providecommand{\eprint}[2][]{\url{#2}}

\bibitem[{\citenamefont{Miskin and Jaeger}(2013)}]{miskin13}
\bibinfo{author}{\bibfnamefont{M.~Z.} \bibnamefont{Miskin}} \bibnamefont{and}
  \bibinfo{author}{\bibfnamefont{H.~M.} \bibnamefont{Jaeger}},
  \bibinfo{journal}{Nature Materials} \textbf{\bibinfo{volume}{12}},
  \bibinfo{pages}{326} (\bibinfo{year}{2013}).

\bibitem[{\citenamefont{Miskin and Jaeger}(2014)}]{miskin14}
\bibinfo{author}{\bibfnamefont{M.~Z.} \bibnamefont{Miskin}} \bibnamefont{and}
  \bibinfo{author}{\bibfnamefont{H.~M.} \bibnamefont{Jaeger}},
  \bibinfo{journal}{Soft Matter} \textbf{\bibinfo{volume}{10}},
  \bibinfo{pages}{3708} (\bibinfo{year}{2014}).

\bibitem[{\citenamefont{Zou et~al.}(2009)\citenamefont{Zou, Cheng, Rivers,
  Jaeger, and Nagel}}]{zou09}
\bibinfo{author}{\bibfnamefont{L.-N.} \bibnamefont{Zou}},
  \bibinfo{author}{\bibfnamefont{X.}~\bibnamefont{Cheng}},
  \bibinfo{author}{\bibfnamefont{M.~L.} \bibnamefont{Rivers}},
  \bibinfo{author}{\bibfnamefont{H.~M.} \bibnamefont{Jaeger}},
  \bibnamefont{and} \bibinfo{author}{\bibfnamefont{S.~R.} \bibnamefont{Nagel}},
  \bibinfo{journal}{Science} \textbf{\bibinfo{volume}{326}},
  \bibinfo{pages}{408} (\bibinfo{year}{2009}).

\bibitem[{\citenamefont{Sacanna et~al.}(2010)\citenamefont{Sacanna, Irvine,
  Chaikin, and Pine}}]{sacanna10}
\bibinfo{author}{\bibfnamefont{S.}~\bibnamefont{Sacanna}},
  \bibinfo{author}{\bibfnamefont{W.~T.~M.} \bibnamefont{Irvine}},
  \bibinfo{author}{\bibfnamefont{P.~M.} \bibnamefont{Chaikin}},
  \bibnamefont{and} \bibinfo{author}{\bibfnamefont{D.~J.} \bibnamefont{Pine}},
  \bibinfo{journal}{Nature} \textbf{\bibinfo{volume}{464}},
  \bibinfo{pages}{575} (\bibinfo{year}{2010}).

\bibitem[{\citenamefont{Brown et~al.}(2012)\citenamefont{Brown, Nasto,
  Athanassiadis, and Jaeger}}]{brown2012}
\bibinfo{author}{\bibfnamefont{E.}~\bibnamefont{Brown}},
  \bibinfo{author}{\bibfnamefont{A.}~\bibnamefont{Nasto}},
  \bibinfo{author}{\bibfnamefont{A.~G.} \bibnamefont{Athanassiadis}},
  \bibnamefont{and} \bibinfo{author}{\bibfnamefont{H.~M.}
  \bibnamefont{Jaeger}}, \bibinfo{journal}{Phys. Rev. Lett.}
  \textbf{\bibinfo{volume}{108}}, \bibinfo{pages}{108302}
  (\bibinfo{year}{2012}).

\bibitem[{\citenamefont{Vutukuri et~al.}(2012)\citenamefont{Vutukuri, Demirors,
  Peng, van Oostrum, Imhof, and van Blaaderen}}]{blaaderen2012}
\bibinfo{author}{\bibfnamefont{H.~R.} \bibnamefont{Vutukuri}},
  \bibinfo{author}{\bibfnamefont{A.~F.} \bibnamefont{Demirors}},
  \bibinfo{author}{\bibfnamefont{B.}~\bibnamefont{Peng}},
  \bibinfo{author}{\bibfnamefont{P.~D.~J.} \bibnamefont{van Oostrum}},
  \bibinfo{author}{\bibfnamefont{A.}~\bibnamefont{Imhof}}, \bibnamefont{and}
  \bibinfo{author}{\bibfnamefont{A.}~\bibnamefont{van Blaaderen}},
  \bibinfo{journal}{Angew. Chem. Int. Ed.} \textbf{\bibinfo{volume}{51}},
  \bibinfo{pages}{11249} (\bibinfo{year}{2012}).

\bibitem[{\citenamefont{Feng et~al.}(2013)\citenamefont{Feng, Dreyfus, Chaikin,
  and Brujic}}]{feng13}
\bibinfo{author}{\bibfnamefont{L.}~\bibnamefont{Feng}},
  \bibinfo{author}{\bibfnamefont{R.}~\bibnamefont{Dreyfus}},
  \bibinfo{author}{\bibfnamefont{P.}~\bibnamefont{Chaikin}}, \bibnamefont{and}
  \bibinfo{author}{\bibfnamefont{J.}~\bibnamefont{Brujic}},
  \bibinfo{journal}{Soft Matter} \textbf{\bibinfo{volume}{9}},
  \bibinfo{pages}{9816} (\bibinfo{year}{2013}).

\bibitem[{\citenamefont{Flory}(1953)}]{flory53}
\bibinfo{author}{\bibfnamefont{P.~J.} \bibnamefont{Flory}},
  \emph{\bibinfo{title}{Principles of Polymer Chemistry}}
  (\bibinfo{publisher}{Cornell University Press}, \bibinfo{address}{Ithaca, New
  York}, \bibinfo{year}{1953}).

\bibitem[{\citenamefont{Karayiannis and
  Laso}(2008{\natexlab{a}})}]{karayiannis08}
\bibinfo{author}{\bibfnamefont{N.~C.} \bibnamefont{Karayiannis}}
  \bibnamefont{and} \bibinfo{author}{\bibfnamefont{M.}~\bibnamefont{Laso}},
  \bibinfo{journal}{Phys. Rev. Lett.} \textbf{\bibinfo{volume}{100}},
  \bibinfo{pages}{050602} (\bibinfo{year}{2008}{\natexlab{a}}).

\bibitem[{\citenamefont{Karayiannis and
  Laso}(2008{\natexlab{b}})}]{karayiannis08b}
\bibinfo{author}{\bibfnamefont{N.~C.} \bibnamefont{Karayiannis}}
  \bibnamefont{and} \bibinfo{author}{\bibfnamefont{M.}~\bibnamefont{Laso}},
  \bibinfo{journal}{Macromolecules} \textbf{\bibinfo{volume}{41}},
  \bibinfo{pages}{1537} (\bibinfo{year}{2008}{\natexlab{b}}).

\bibitem[{\citenamefont{Foteinopoulou et~al.}(2008)\citenamefont{Foteinopoulou,
  Karayiannis, Laso, Kr{\"o}ger, and Mansfield}}]{foteinopoulou08}
\bibinfo{author}{\bibfnamefont{K.}~\bibnamefont{Foteinopoulou}},
  \bibinfo{author}{\bibfnamefont{N.~C.} \bibnamefont{Karayiannis}},
  \bibinfo{author}{\bibfnamefont{M.}~\bibnamefont{Laso}},
  \bibinfo{author}{\bibfnamefont{M.}~\bibnamefont{Kr{\"o}ger}},
  \bibnamefont{and} \bibinfo{author}{\bibfnamefont{M.~L.}
  \bibnamefont{Mansfield}}, \bibinfo{journal}{Phys. Rev. Lett.}
  \textbf{\bibinfo{volume}{101}}, \bibinfo{pages}{265702}
  (\bibinfo{year}{2008}).

\bibitem[{\citenamefont{Karayiannis
  et~al.}(2009{\natexlab{a}})\citenamefont{Karayiannis, Foteinopoulou, and
  Laso}}]{karayiannis09}
\bibinfo{author}{\bibfnamefont{N.~C.} \bibnamefont{Karayiannis}},
  \bibinfo{author}{\bibfnamefont{K.}~\bibnamefont{Foteinopoulou}},
  \bibnamefont{and} \bibinfo{author}{\bibfnamefont{M.}~\bibnamefont{Laso}},
  \bibinfo{journal}{Phys. Rev. Lett.} \textbf{\bibinfo{volume}{103}},
  \bibinfo{pages}{045703} (\bibinfo{year}{2009}{\natexlab{a}}).

\bibitem[{\citenamefont{Karayiannis
  et~al.}(2009{\natexlab{b}})\citenamefont{Karayiannis, Foteinopoulou, and
  Laso}}]{karayiannis09b}
\bibinfo{author}{\bibfnamefont{N.~C.} \bibnamefont{Karayiannis}},
  \bibinfo{author}{\bibfnamefont{K.}~\bibnamefont{Foteinopoulou}},
  \bibnamefont{and} \bibinfo{author}{\bibfnamefont{M.}~\bibnamefont{Laso}},
  \bibinfo{journal}{J. Chem. Phys.} \textbf{\bibinfo{volume}{130}},
  \bibinfo{pages}{164908} (\bibinfo{year}{2009}{\natexlab{b}}).

\bibitem[{\citenamefont{Karayiannis et~al.}(2010)\citenamefont{Karayiannis,
  Foteinopoulou, Abrams, and Laso}}]{karayiannis10}
\bibinfo{author}{\bibfnamefont{N.~C.} \bibnamefont{Karayiannis}},
  \bibinfo{author}{\bibfnamefont{K.}~\bibnamefont{Foteinopoulou}},
  \bibinfo{author}{\bibfnamefont{C.~F.} \bibnamefont{Abrams}},
  \bibnamefont{and} \bibinfo{author}{\bibfnamefont{M.}~\bibnamefont{Laso}},
  \bibinfo{journal}{Soft Matter} \textbf{\bibinfo{volume}{6}},
  \bibinfo{pages}{2160} (\bibinfo{year}{2010}).

\bibitem[{\citenamefont{Karayiannis
  et~al.}(2009{\natexlab{c}})\citenamefont{Karayiannis, Foteinopoulou, and
  Laso}}]{karayiannis10b}
\bibinfo{author}{\bibfnamefont{N.~C.} \bibnamefont{Karayiannis}},
  \bibinfo{author}{\bibfnamefont{K.}~\bibnamefont{Foteinopoulou}},
  \bibnamefont{and} \bibinfo{author}{\bibfnamefont{M.}~\bibnamefont{Laso}},
  \bibinfo{journal}{Phys. Rev. E} \textbf{\bibinfo{volume}{80}},
  \bibinfo{pages}{011307} (\bibinfo{year}{2009}{\natexlab{c}}).

\bibitem[{\citenamefont{Lopatina et~al.}(2011)\citenamefont{Lopatina, {Olson
  Reichhardt}, and Reichhardt}}]{reichhardt11}
\bibinfo{author}{\bibfnamefont{L.~M.} \bibnamefont{Lopatina}},
  \bibinfo{author}{\bibfnamefont{C.~J.} \bibnamefont{{Olson Reichhardt}}},
  \bibnamefont{and}
  \bibinfo{author}{\bibfnamefont{C.}~\bibnamefont{Reichhardt}},
  \bibinfo{journal}{Phys. Rev. E} \textbf{\bibinfo{volume}{84}},
  \bibinfo{pages}{011303} (\bibinfo{year}{2011}).

\bibitem[{\citenamefont{Karayiannis et~al.}(2013)\citenamefont{Karayiannis,
  Foteinopoulou, and Laso}}]{karayiannis13b}
\bibinfo{author}{\bibfnamefont{N.~C.} \bibnamefont{Karayiannis}},
  \bibinfo{author}{\bibfnamefont{K.}~\bibnamefont{Foteinopoulou}},
  \bibnamefont{and} \bibinfo{author}{\bibfnamefont{M.}~\bibnamefont{Laso}},
  \bibinfo{journal}{Philos. Mag.} \textbf{\bibinfo{volume}{93}},
  \bibinfo{pages}{4108} (\bibinfo{year}{2013}).

\bibitem[{\citenamefont{Ni and Dijkstra}(2013)}]{ni13}
\bibinfo{author}{\bibfnamefont{R.}~\bibnamefont{Ni}} \bibnamefont{and}
  \bibinfo{author}{\bibfnamefont{M.}~\bibnamefont{Dijkstra}},
  \bibinfo{journal}{Soft Matter} \textbf{\bibinfo{volume}{9}},
  \bibinfo{pages}{365} (\bibinfo{year}{2013}).

\bibitem[{\citenamefont{Hoy and Karayiannis}(2013)}]{hoy13}
\bibinfo{author}{\bibfnamefont{R.~S.} \bibnamefont{Hoy}} \bibnamefont{and}
  \bibinfo{author}{\bibfnamefont{N.~C.} \bibnamefont{Karayiannis}},
  \bibinfo{journal}{Phys. Rev. E} \textbf{\bibinfo{volume}{88}},
  \bibinfo{pages}{012601} (\bibinfo{year}{2013}).

\bibitem[{\citenamefont{Plimpton}(1995)}]{plimpton95}
\bibinfo{author}{\bibfnamefont{S.}~\bibnamefont{Plimpton}},
  \bibinfo{journal}{J. Comp. Phys.} \textbf{\bibinfo{volume}{117}},
  \bibinfo{pages}{1} (\bibinfo{year}{1995}).

\bibitem[{pth()}]{pthres}
\bibinfo{note}{Choosing a lower (higher) value of $P_{thres}$ lowers (raises)
  $\phi_J(\theta_0)$ and $Z_{nc}[\phi=\phi_J(\theta_0)]$, but does not
  qualitatively change any of the results presented herein.}

\bibitem[{\citenamefont{van Hecke}(2009)}]{vanHecke09}
\bibinfo{author}{\bibfnamefont{M.}~\bibnamefont{van Hecke}},
  \bibinfo{journal}{J. Phys. Cond. Matt.} \textbf{\bibinfo{volume}{22}},
  \bibinfo{pages}{033101} (\bibinfo{year}{2009}).

\bibitem[{\citenamefont{O'Hern et~al.}(2003)\citenamefont{O'Hern, Silbert, Liu,
  and Nagel}}]{ohern02}
\bibinfo{author}{\bibfnamefont{C.~S.} \bibnamefont{O'Hern}},
  \bibinfo{author}{\bibfnamefont{L.~E.} \bibnamefont{Silbert}},
  \bibinfo{author}{\bibfnamefont{A.~J.} \bibnamefont{Liu}}, \bibnamefont{and}
  \bibinfo{author}{\bibfnamefont{S.~R.} \bibnamefont{Nagel}},
  \bibinfo{journal}{Phys. Rev. E} \textbf{\bibinfo{volume}{68}},
  \bibinfo{pages}{011306} (\bibinfo{year}{2003}).

\bibitem[{\citenamefont{Donev et~al.}(2004)\citenamefont{Donev, Stillinger,
  Chaikin, and Torquato}}]{donev04}
\bibinfo{author}{\bibfnamefont{A.}~\bibnamefont{Donev}},
  \bibinfo{author}{\bibfnamefont{F.~H.} \bibnamefont{Stillinger}},
  \bibinfo{author}{\bibfnamefont{P.~M.} \bibnamefont{Chaikin}},
  \bibnamefont{and} \bibinfo{author}{\bibfnamefont{S.}~\bibnamefont{Torquato}},
  \bibinfo{journal}{Phys. Rev. Lett.} \textbf{\bibinfo{volume}{92}},
  \bibinfo{pages}{255506} (\bibinfo{year}{2004}).

\bibitem[{\citenamefont{Mailman et~al.}(2009)\citenamefont{Mailman, Schreck,
  O'Hern, and Chakraborty}}]{mailman09}
\bibinfo{author}{\bibfnamefont{M.}~\bibnamefont{Mailman}},
  \bibinfo{author}{\bibfnamefont{C.~F.} \bibnamefont{Schreck}},
  \bibinfo{author}{\bibfnamefont{C.~S.} \bibnamefont{O'Hern}},
  \bibnamefont{and}
  \bibinfo{author}{\bibfnamefont{B.}~\bibnamefont{Chakraborty}},
  \bibinfo{journal}{Phys. Rev. Lett.} \textbf{\bibinfo{volume}{102}},
  \bibinfo{pages}{255501} (\bibinfo{year}{2009}).

\bibitem[{\citenamefont{Kyrylyuk and Philipse}(2011)}]{kyrylyuk11}
\bibinfo{author}{\bibfnamefont{A.~V.} \bibnamefont{Kyrylyuk}} \bibnamefont{and}
  \bibinfo{author}{\bibfnamefont{A.~P.} \bibnamefont{Philipse}},
  \bibinfo{journal}{Phys. Status Solidi A} \textbf{\bibinfo{volume}{208}},
  \bibinfo{pages}{2299} (\bibinfo{year}{2011}).

\bibitem[{non()}]{nonsmooth}
\bibinfo{note}{The nonsmooth behavior of $\phi_J(\theta_0)$ for $N=100$ chains
  in the range $5^{\circ} \lesssim \theta_0 \lesssim 15^{\circ}$ (Fig.\
  \ref{fig:phiJ}(a)) appears to be a finite-system-size effect caused by chains
  interacting with their periodic images.}

\bibitem[{phi()}]{phijmon}
\bibinfo{note}{Previous studies
  \cite{karayiannis08,karayiannis09b,karayiannis10b} that found $\phi_J^{flex}$
  equal to the value for monomers ($\phi_J^{mon}=.636$ \cite{ohern02}) used
  topology-changing algorithms \cite{karayiannis08b} that eliminated any
  contribution of chain connectivity to the jamming process.}

\bibitem[{\citenamefont{Broedersz et~al.}(2011)\citenamefont{Broedersz, Mao,
  Lubensky, and Mackintosh}}]{broedersz11}
\bibinfo{author}{\bibfnamefont{C.~P.} \bibnamefont{Broedersz}},
  \bibinfo{author}{\bibfnamefont{X.}~\bibnamefont{Mao}},
  \bibinfo{author}{\bibfnamefont{T.~C.} \bibnamefont{Lubensky}},
  \bibnamefont{and} \bibinfo{author}{\bibfnamefont{F.~C.}
  \bibnamefont{Mackintosh}}, \bibinfo{journal}{Nature Phys.}
  \textbf{\bibinfo{volume}{7}}, \bibinfo{pages}{983} (\bibinfo{year}{2011}).

\bibitem[{\citenamefont{de~Gennes}(1979)}]{degennes79}
\bibinfo{author}{\bibfnamefont{P.~G.} \bibnamefont{de~Gennes}},
  \emph{\bibinfo{title}{Scaling Concepts in Polymer Physics}}
  (\bibinfo{publisher}{Cornell University Press (Ithaca)},
  \bibinfo{year}{1979}).

\bibitem[{pol()}]{polytet}
\bibinfo{note}{The other peaks in $P(\psi)$ for flexible polymers reported in
  Ref.\ \cite{karayiannis09b} are missing here, presumably because the
  topology-changing moves \cite{karayiannis08b} used therein, that negate chain
  uncrossability, alter the character of chain collapse.}

\bibitem[{\citenamefont{Anikeenko and Medvedev}(2007)}]{anikeenko07}
\bibinfo{author}{\bibfnamefont{A.~V.} \bibnamefont{Anikeenko}}
  \bibnamefont{and} \bibinfo{author}{\bibfnamefont{N.~N.}
  \bibnamefont{Medvedev}}, \bibinfo{journal}{Phys. Rev. Lett.}
  \textbf{\bibinfo{volume}{98}}, \bibinfo{pages}{235504}
  (\bibinfo{year}{2007}).

\bibitem[{\citenamefont{Strobl}(2007)}]{strobl07}
\bibinfo{author}{\bibfnamefont{G.}~\bibnamefont{Strobl}},
  \emph{\bibinfo{title}{The Physics of Polymers}}
  (\bibinfo{publisher}{Springer}, \bibinfo{year}{2007}).

\bibitem[{\citenamefont{Ballauf}(1989)}]{ballauff89}
\bibinfo{author}{\bibfnamefont{M.}~\bibnamefont{Ballauf}},
  \bibinfo{journal}{Angew. Chem.} \textbf{\bibinfo{volume}{28}},
  \bibinfo{pages}{253} (\bibinfo{year}{1989}).

\end{thebibliography}

\end{document}